# Realist Analysis of Six Controversial Quantum Issues

Art Hobson[1]

**Abstract.**  This paper presents a philosophically realistic analysis of quantization, field-particle duality, superposition, entanglement, nonlocality, and measurement.  These are logically related:  Realistically understanding measurement depends on realistically understanding superposition, entanglement, and nonlocality; understanding these three depends on understanding field-particle duality and quantization.  This paper resolves all six, based on a realistic view of standard quantum physics.  It concludes that, for these issues, standard quantum physics is consistent with scientific practice since Copernicus:  Nature exists on its own and science's goal is to understand its operating principles, which are independent of humans.  Quantum theory need not be regarded as merely the study of what humans can know about the microscopic world, but can instead view it as the study of real quanta such as electrons, photons, and atoms.  This position has long been argued by Mario Bunge.

**Introduction**

Quantum physics (QP) is in a scandalous state (van Kampen 2008).  Although founded at the dawn of the *preceding* century, and although arguably the most wide-ranging, highly accurate, and economically rewarding scientific theory of all time (Hobson 2017, pp.7-10), its fundamentals remain in disarray.  In particular, there is little consensus about wave-particle duality, superposition, entanglement, nonlocality, and measurement.

This disarray has spawned an astonishing assortment of interpretations and alterations of the theory.  A quantum foundations conference polled its 33 expert attendees about their favorite interpretation (Schlosshauer 2013).  Nonrealistic interpretations, which claim that QP describes only our knowledge of the microworld rather than the microworld itself, gathered 24 votes.  These sub-divided as 14 for the Copenhagen interpretation, 8 for the view that QP is about "information," and 2 for "Bayesianism" according to which quantum states represent *personal* degrees of belief.  Six chose the many worlds interpretation, which is realistic in the sense that the trillions of universes supposedly created every time one photon strikes a light-sensitive surface are presumed to be real.  The remainder chose "other" or "none."  Remarkably missing from the listed options was "standard quantum physics realistically interpreted."

It's not sufficiently recognized that this quantum confusion has spawned a serious social threat.  Pseudoscience and related fantasies are rampant, especially in my country (Andersen 2017).  Earth cries out for rationality and scientific literacy (Sagan 1995), yet quantum-inspired pseudoscience has become a threat to both.  A pseudoscientific movie, *What the bleep do we know,* won several film awards and grossed $10 million in 2004; it argues that we create our own reality through consciousness and QP, and features physicists saying things like "The material world around us is nothing but possible movements of consciousness" (Shermer 2005).  A non-technical physics literacy textbook being taught in high schools and universities bears the title *Quantum*

---

[1] Department of Physics, University of Arkansas, Fayetteville, USA.  Email: ahobson@uark.edu

*Enigma: Physics Encounters Consciousness* (Rosenblum & Kuttner 2006). According to the book's dust cover, "Every interpretation of QP encounters consciousness." It's striking that Stenger (1997) and Shermer (2005) have, by coincidence, the same title: *Quantum Quackery.*[2]

This essay provides a philosophically realistic analysis of six phenomena central to our current confusion: quantization, field-particle duality, superposition, entanglement, nonlocality, and measurement. As will be shown, these are logically related: Understanding measurement depends on properly understanding nonlocality, entanglement, and superposition; understanding these three depends on properly understanding field-particle duality and quantization. All six will be resolved, based on a realistic interpretation of standard quantum physics. The analysis is internally consistent as well as consistent with the relevant experimental facts. Thus, at least for these issues, QP concurs with the scientific paradigm as it has been known since Copernicus: nature exists on its own and science's goal is to understand nature's operating principles, which are independent of humans.

Non-realistic interpretations generally regard QP to be merely the study of "what we can say" about the microworld rather than the study of the actual microworld. As Niels Bohr reportedly stated:

> There is no quantum world. There is only an abstract quantum physical description. It is wrong to think that the task of physics is to find out how nature is. Physics concerns what we can say about nature (quoted in French 1985, p.305).

At least with regard to the analyzed issues, this non-realistic view need not be accepted. It is possible, instead, to regard QP as the study of the general principles of the microworld itself, just as fluid dynamics is the study of fluids and geology is the study of Earth's structure. This conclusion concurs with Mario Bunge's lifetime work and further verifies his conclusions (Bunge 1967, 2003a, 2003b, 2012, Mahner 2001).

**The Quantum**

The quantum entered history on 14 December 1900 at a meeting of the Deutsche Physikalische Gesellschaft in Berlin when Max Planck first publicly wrote his famous $E = hf$. We view this today as the equation for the energy (in joules) of a photon having frequency $f$ (in Hertz) where $h = 6.6 \times 10^{-34}$ joule-seconds is Planck's constant to two figures.

It wasn't always thought of that way, certainly not by Planck in 1900. For Planck, $E = hf$ represented the best possible fudge factor. He, like other physicists, had been searching for a theoretical explanation of the electromagnetic (EM) energy radiated at each frequency by a "black body" such as a small opening in an oven. Previous explanations predicted that the energy radiated at short wavelengths (high frequencies) was so enormous it should blind us whenever we look at a fire (Tegmark & Wheeler 2001)! Planck assumed correctly that radiation comes from vibrations of atoms such as those on the inner walls of the oven. Unable to solve the ensuing mathematics, he resorted to the numerical trick of assuming the amount of energy emitted by an atom could not range over a continuum of values but was instead restricted to a discrete set of values 0, $\varepsilon$, $2\varepsilon$, $3\varepsilon$, etc., where the increment $\varepsilon$ is a small number. Today, we would say he assumed the amount of energy emitted to be "quantized." His plan was to calculate the radiation formula in this simpler approximation, and then find the real result by allowing $\varepsilon$ to approach 0 since, as everybody knew, energies can always vary continuously. But, as happens frequently in quantum history, what "everybody knew" turned out to be wrong. When Planck allowed ε to approach 0, he recovered the bizarre result that others had found. It turned out the only way to get agreement with experiment was to allow ε to be the *non-zero* frequency-dependent number *hf*.

---

[2] For further discussion, see Hobson 2013 and 2017, pp.12-13.

Planck had happened upon a principle deeper and stranger than he could have known: microscopic energies range over a discrete, rather than continuous, set of possible values.

This is the source of quantum jumps. Here's why. If the energy emitted by an atom must be an integral multiple of a positive number ε, then an atom must emit energy in *instantaneous bursts,* rather than continuously over time. That is, the least energy an atom can radiate is ε joules, and *this energy must be emitted at a single instant*. This is because, if ε joules are emitted continuously over some positive time $\tau > 0$, then some fraction of ε joules must be emitted in any shorter time, but *Planck's rule does not allow this*. Thus, there are physical "processes" or, better, "events," involving positive amounts of energy, that happen in "zero"[3] time. Today we call them quantum jumps. Discontinuity and quantum jumps were baked into QP from the beginning, and we should not be disconcerted when they re-appear in atoms, nonlocal phenomena, measurement, etc.

The greatest works of art are surprising. A Beethoven symphony's change of key, a Picasso painting's fractured face, are entirely unexpected yet, once we experience them, we realize they are perfect. The quantum is perhaps nature's ultimate stroke of genius. Quantum physics is the study of these quanta.

Little attention was paid to Planck's idea until 1905, when Albert Einstein reasoned that, if EM energy is emitted in small (in energy but not necessarily in spatial extent) lumps, then it probably travels through space in small lumps; he employed this notion to explain the emission of electrons from metal surfaces struck by radiation of sufficiently high frequency. Classical EM theory was unable to explain this in terms of smoothly extended EM fields, but Planck's energy lumps did the trick. Later, the lumps came to be called "photons." Einstein's contribution was crucial, for it extended the notion of the quantum as a finite small increment of energy to the notion of the quantum as a physical object: an energetically small and highly unified lump of energy. Photons, electrons, and quarks are examples of simple, or fundamental, quanta. Atoms and molecules are also quanta, but they are made of simpler quanta "entangled" into a highly unified "compound" quantum. All quanta are made of "quantized" (i.e. obeying the principles of QP) amounts of various sorts of energy. For photons, electrons, and quarks, the energies arise, respectively, from the EM field, electron-positron field, and strong field, three "quantum fields" that pervade the universe.

Einstein was the first to notice that quantum discontinuity has nonlocal implications. At the Firth Solvay Conference in Brussels in 1927, shortly after Heisenberg and Schrodinger invented quantum theory, Einstein asked the gathering to consider a single electron passing through a small hole in a partition. Schrodinger's equation predicts that its wave function Ψ, after passing through the hole, spreads out broadly over a distant viewing screen which Einstein assumed was spherical so Ψ would reach all of it simultaneously. Yet the electron impacts at only a small place on the screen. Einstein's written version states:

> The scattered wave moving towards [the viewing screen] does not present any preferred
> direction. If psi-squared were simply considered as the probability that a definite particle is situated at a certain place at a definite instant, it might happen that *one and the same* elementary process would act *at two or more* places of the screen. But the interpretation, according to which psi-squared expresses the probability that *this* particle is situated at a certain place, presupposes a very particular mechanism of action at a distance (Gilder 2008, p.374).

---

[3] But taking into consideration possible fundamental limitations such as the Planck time.

Einstein's concern was that, if the impact appears at some point $x$, then the impact must also *not* appear at other points $y$, so the status of such other points must *instantly* switch from "possible impact point" to "impossible impact point." Instantaneous correlations must therefore exist between $x$ and $y$, and special relativity does not seem to allow this. This was five years prior to the first broad analysis of measurement (von Neumann 1932) and thirty-seven years prior to the earliest real understanding of nonlocality (Bell 1964).

Planck had practical reasons for inventing his fudge factor, but for nature it's not a fudge factor. Why the quantum? It's a question famously asked by John Wheeler (Barrow et al. 2004). My suggestion would be that nature prefers countable sets, or perhaps even finite sets, over continuous and therefore uncountable sets. There are many more numbers along any single interval of the continuous real line than there are integers. If the universe is constructed out of countable quanta, reality becomes *much* simpler. This suggests that the universe may even be constructed from a *finite* number of quanta, which would be far simpler still.

The countability of quanta might suggest that they are particles. But as will be demonstrated, quanta are not particles and are not necessarily even tiny.

**Is Reality Made of Fields, Particles, or Both?**

The notion that everything is made of tiny particles runs through scientific history. Early Greeks such as Democritus of Abdera perceived reality to be made of small material particles moving in empty space. Newton's physics begins from particles obeying Newton's laws, so the motion of complex objects follows from that of their particles. Indeed, Newton writes:

> All these things being considered, it seems probable to me that God in the beginning formed matter in solid, massy, hard, impenetrable, movable particles, of such sizes and figures, and with such other properties, and in such proportion to space, as most conduced to the end for which he formed them (Newton 1998/1704).

The particle notion runs deep. Schrodinger's equation for, say, a moving electron is clearly a "field equation" for a scalar (i.e. number-valued rather than vector-valued) field $\Psi(x, y, z, t)$ entirely analogous to Maxwell's equations for the vector EM field $\mathbf{E}(x, y, z, t)$, $\mathbf{B}(x, y, z, t)$. Nevertheless, the QP founders retained the Newtonian language, speaking consistently of quantum "mechanics" and quantum "particles." Physicists continue to apply the term "particle" to essentially every quantum object, including even the Higgs field which is clearly a universe-filling quantized field. Because language so shapes our perception of reality, I doubt we will transcend our quantum confusion until we adopt more appropriate words. Electrons, photons, and other quanta are not "particles."

The question of fields versus particles is crucial because, once one adopts the particle misconception, most other issues become unfathomable. For example, if quanta are particles separated by empty space, distant nonlocal connections become incomprehensible.

The best example is still the double-slit experiment.[4] The set-up is a beam of monochromatic light or mono-energetic electrons passing through a pair of parallel narrow vertical slits cut into an opaque partition, with a viewing screen beyond the slits. This experiment with light was performed in 1801 by Thomas Young, who found that the viewing screen displayed an interference pattern: many vertical bands of light separated by dark bands. Physicists correctly inferred that light is a wave in a field later identified as the universal EM field. The interference arises from the intersection of two light waves, one from each slit, with "constructive interference" (the light bands) arising where crests meet crests and "destructive interference" (the dark bands) arising where crests meet valleys. Such waves indicate an underlying field that "carries" the waves. Thus, light is a wave in the universal EM field, much as ocean waves are waves in the water. This was the consensus among 19th-century physicists, and it's correct today.

The analogous experiment with an electron beam was performed for the first time by Claus Jonsson in 1961. The long-predicted result was an interference pattern just like Young's result! The only significant difference: Jonsson's interference bands were much smaller because typical electron wavelengths are smaller than photon wavelengths.

But most of us were taught that light is a spatially extended nonmaterial wave while electrons are tiny material particles orbiting within atoms. How can electrons be similar to light?

Here's how. In either Young's or Jonsson's experiment, one can dim the beam so extremely that only a single quantum (photon or electron) comes through at a time. Each quantum makes a tiny flash somewhere on the viewing screen. After 10 to 100 quanta have impacted, and assuming each flash persists indefinitely, we see 10 to 100 flashes distributed apparently randomly over the entire screen. But by the time 1,000 quanta have impacted the screen, we begin to see an interference pattern forming from the distribution of flashes, the way a pointillist painting forms from tiny dots. With more impacts, it becomes clear that both interference patterns result from large numbers of small impacts. The similarity of the two experiments defies the notion that the first arises from waves and the second from particles.

Either experiment can be performed using large numbers of identically-prepared photons or electrons. The interference pattern that spreads over the entire screen then demonstrates quantum randomness: Identical preparations but a variety of outcomes. However, impacts are not entirely random but tend to cluster in regions of constructive interference. Since preparations are identical, *the entire pattern must be carried, i.e. "known," by each quantum*.

This suggests trying the experiment with only one slit open. The interference bands then vanish, replaced by a broad swath of tiny impacts distributed randomly all over the screen.[5] This is perplexing if quanta are particles. If a quantum is a small particle, then by definition it can come through only one slit regardless of whether the other slit is open. How does it "know" whether the other, relatively distant, slit is open? The experiment has been done not only with photons and electrons, but also with neutrons, atoms and molecules of all sorts, and there are even plans to use viruses. It's hard to imagine any long-range force that could inform these quanta as to whether the other slit is open or closed. In fact the forthright particles advocate Richard Feynman said "you will get down the drain into a blind alley" if you think too hard about this (Feynman 1965).

According to the field view, every photon or electron is a wave rippling through a universe-filling field, namely the "quantized" (i.e. obeying quantum rather than classical rules) EM field or positron-electron field. Fields, because they naturally spread, resolve the question that got Feynman down the drain: Each photon or electron has sufficient spatial extension to come through both slits, encoding the double-slit or single-slit pattern. As Paul Dirac put it, "Each photon ...interferes only with itself" (Dirac 1958).

---

[4] The remainder of this section is based on (Hobson 2013), which should be consulted for references and details.

[5] This assumes the slit's width is smaller than the wavelength, so the light passing through the slit spreads out into a single broad diffraction band without side fringes.

So fields explain the interference, but what about the particle-like impact points? This is the "measurement problem," explained more fully below. Briefly, the quantum approaching the screen extends over the entire interference pattern, and instantaneously "collapses" to microscopic size upon impact. This happens because the quantum "entangles" with the screen's atoms, instantaneously "localizing" the quantum. Most physicists would not phrase it this way. The standard words are that the quantum's "wave function" describes the entire pattern, and the "wave function" then collapses upon impact. But this is mincing words. Why would one invent a new object, the "wavefunction," to explain measurements? The quantum and its wavefunction are not separate objects. The quantum is its wavefunction. Fields are all there is.

This is a good place to clarify a confusion that annoyed John Bell and many others. Bell complained that QP seems "exclusively concerned about 'results of measurement,' and has nothing to say about anything else." Who. he asked, qualifies as the measurer? Is it every living creature? Or must it be a physicist? "The aim remains to understand the world. To restrict quantum mechanics to be exclusively about piddling laboratory operations is to betray the great enterprise. A serious formulation will not exclude the big world outside the laboratory" (Bell 1990).

It's hard to understand why this point wasn't forcefully stated long before 1990. The answer is not difficult. Every quantum measurement involves a measuring device that detects a quantum object and records a macroscopic mark. This happens not only in laboratories, it happens all the time in nature as for example when a cosmic ray strikes and moves a sand grain on Mars. A "quantum measurement" is simply any process by which a quantum phenomenon causes a macroscopic change. It has nothing necessarily to do with humans.

On a similar note, the "Born rule" connecting a quantum's state (or wavefunction) $\Psi$ with the macroscopic world is usually stated this way: "The probability density for finding the quantum at some point $x$ upon measurement is $\Psi$-mod-squared evaluated at $x$." The word "finding" is inappropriate because it suggests a human who does the finding, and it suggests the measured quantum is "at" some single point $x$ and is therefore a particle. The quoted statement should be replaced by: "If a quantum interacts with another object, the probability density for the interaction to occur at $x$ is $\Psi$-mod-squared evaluated at $x$."

Further linguistic points: Planck's energy lumps are "quanta of the EM field." There are other kinds of quanta: Electrons are quanta of the electron-positron field, quarks are quanta of the strong field, etc. "Quantum physics" is the study of fields that are bundled into highly unified lumps called "quanta."

Standard non-relativistic QP based on Schrodinger's equation is not the best basis for analyzing field-particle duality, because spontaneous quantum energy fluctuations plus relativity's principle of mass-energy equivalence entail that quanta can be created or destroyed. Relativistic QP was invented to deal with such possibilities. But it isn't easy to fit QP into a relativistic framework. In fact, rigorous theorems show that particles cannot exist in a universe that obeys both QP's so-called "unitary" time development and special relativity. The reason is that, even if we generously define a particle as any entity that is "localized" in the sense that it is contained, with 100% probability, within *some* region of finite volume, particles are impossible.

Here's why. If such a particle exists at some specific time such as t=0, then unitary evolution implies it must *instantly* expand and have a positive probability of interacting an arbitrarily large distance away from its initial localization region at any positive time t > 0. But such instantaneous expansion implies that a particle on Earth at t=0 could be detected on the moon an arbitrarily short time later, and this violates special relativity's ban on superluminal transport.

Thus, even under a broadly inclusive definition of "particle," quantum particles contradict special relativity. Fields, which fill all space, do not. Louis de Broglie put it bluntly in his 1924 PhD dissertation (a dissertation that delighted Einstein):

> The energy of an electron is spread over all space with a strong concentration in a very small region. That which makes an electron an atom of energy is not its small volume that it occupies in space, I repeat it occupies all space, but the fact that it is undividable, that it constitutes a unit (Baggott 2011,p.38).

Quantum particles have many other conflicts with relativity. In fact, the only known version of relativistic quantum theory is quantum field theory which is, as its name implies, a theory of fields, not particles (Eberhard & Ross 1989). The Standard Model of high-energy physics, for example, is two quantum field theories known as the electroweak theory and quantum chromodynamics. Non-relativistic quantum theory based on Schrodinger's equation can also be cast as a quantum field theory.

It's well known that quantum field theories predict "vacuum states" having energy but no quanta. This is for the same reason (namely the uncertainty principle) that a mechanical oscillator obeying QP cannot be at rest but must instead have a ground state energy of *hf/2*. The quantum vacuum is quite real: It's the source of exquisitely predicted and verified phenomena such as the Lamb shift, the Casimir effect, and the electron's anomalous magnetic moment. Furthermore, excited states of atoms and other systems could not spontaneously transition to ground states without the quantum vacuum. This quantum vacuum is embarrassing for particle theories. If the universe is made of particles, then what is it that has this energy in the state that has no particles?

Another example of this embarrassment is the Unruh effect. Quantum field theory predicts that an accelerating observer in vacuum observes quanta while a non-accelerating observer of the same vacuum observes no quanta. If particles form the basic reality, how can they be present for the first observer and absent for a second observer of the same region? If fields are basic, this conundrum resolves itself: The first observer's acceleration promotes the second observer's vacuum fluctuations into higher-energy excitations.

Single-quantum nonlocality provides further evidence that the vacuum has properties arising from real fields. In fact, Einstein's remark at the 1927 Solvay Conference, quoted earlier, suggests individual quanta have nonlocal properties, and experimental and theoretical work since 1991 has shown that such properties arise from a single quantum's entanglement with the quantum vacuum (Tan 1991).

In summary, the particle view encounters contradictions at every turn, but everything is consistent under an all-fields view. Particle-like phenomena arise from the discrete, countable nature of quanta, and from a quantum's partial localization upon detection. Picture a single photon, for example, as a spatially-extended cloud big enough to pass through both slits and then to extend over its entire interference region, which collapses into an atom-sized region upon impact.

**Superposition**

Since quanta are waves in fields, it's not surprising that they obey a "superposition principle": If a quantum can be in any one of several different states, then it can be in all of them simultaneously.

Similarly, two water waves can exist simultaneously in the same pond and can even pass through each other. This doesn't seem odd for water waves, which are just shapes of the pond's surface, but it seems odd for, say, electrons and atoms. It entails that quanta can be in two places simultaneously.

The experiment diagrammed in Figure 1 furnishes a perfect example. The figure shows an interference measuring device or "interferometer." Imagine a monochromatic light beam entering at the lower left. The interferometer comprises two "optical paths" along which the beam can travel. BS1 is a "beam splitter," a device that reflects half the beam upward into path 1 and transmits the other half rightward into path 2. M1 and M2 are mirrors that bring the two beams back together so they cross at a second removable beam splitter BS2. D1 and D2 are light detectors, and $\phi_1$ and $\phi_2$ are "phase shifters" that can lengthen path 1 and path 2, respectively, by any amount up to one wavelength of the light.

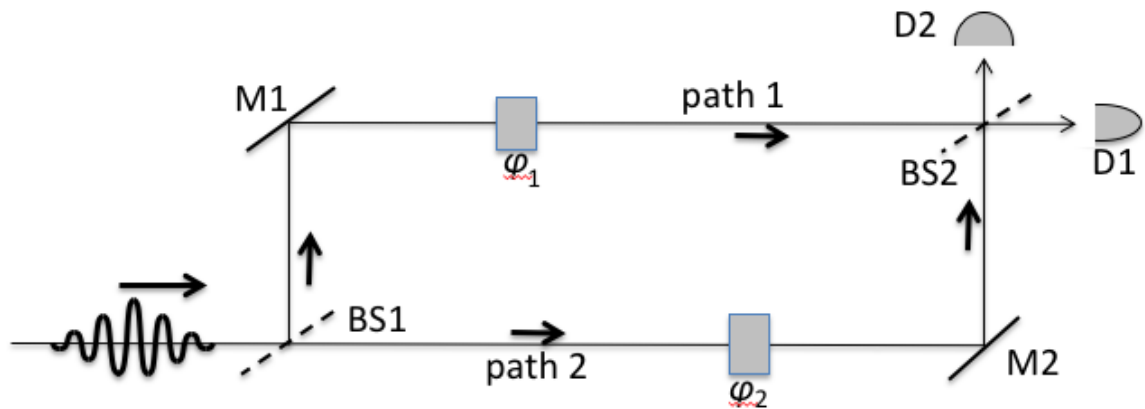

Figure 1. A Mach-Zehnder interferometer.

If we remove BS2 and send a light beam through, half the light goes to D1 and half to D2 independently of how we set the phase shifters: evidence that the two half-beams simply pass through each other at the crossing point.

With BS2 inserted, half the light in each path goes to each detector, so the two paths mix together. The amount of light going to each detector varies as the path lengths change. At some phase settings, all the light goes to D1. This demonstrates interference: constructive at D1 and destructive at D2. If we then lengthen, say, path 1 slightly, a fraction of the light goes to D2. Lengthening path 1 by a quarter wavelength results in half the light going to each detector. Lengthening by a half wavelength results in all the light going to D2. The two paths are clearly interfering at the detectors, verifying that light is a wave.

If we dim the beam sufficiently, quantum effects must show up at some point. What happens when a single photon encounters BS1? Planck's rule entails that the photon cannot split, with half of it following each path. We discover, experimentally, that the entire photon instead reflects *and* transmits. It follows both paths! This sounds crazy. How do we know this?

Here's how. When a single photon passes through the interferometer with BS2 absent we find after many "trials" that the entire photon always impacts either D1 or D2, never both, verifying the photon to be a single unified object that doesn't split. The impacts occur with 50-50 random statistics at D1 and D2. It's interesting and significant that the randomness is *perfect*. For example, with a long enough series of trials we find the perfect fraction (1 part in $2^{10}$) of ten D1s in a row.

Quantum indeterminacy is more perfectly random than any classical device such as a roulette wheel or coin flips (Hobson 2017).

Continuing the explanation: Inserting BS2 and varying the path lengths reveals a surprising result: The statistics at D1/D2 vary depending on the *difference between* the two path lengths, as shown in Figure 2, so the measured outcome responds equally to phase alterations of *either* path. The simplest conclusion is that each photon goes both ways; it is "superposed" along both paths. When the two "copies" of the photon pass through BS2, each copy again goes both ways so the two paths mix and interfere at either detector: If crest meets crest at D1, interference is constructive and we detect a wave (a photon). If crest meets valley, we detect no photon at D1. Figure 2 shows how the probabilities of detecting the photon at D1 vary with the difference between the two path lengths.

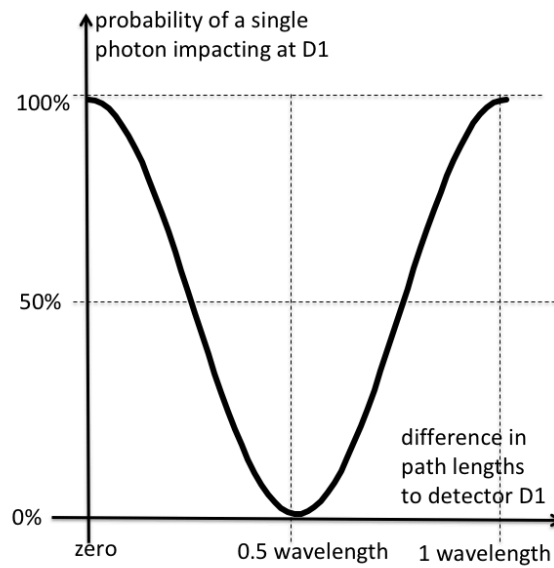

Figure 2. Outcome of the experiment of Figure 1, evidence that each photon travels both paths.

If photons are particles, similar to peas only smaller, superposition is remarkable. But if a photon is an extended real lump of field, it's expected: The field simply spreads along both paths.

Interferometer experiments have also been performed with material quanta such as atoms, with the same result. Atoms, too, are fields.

Gui-Lu Long and colleagues recently performed an interferometer experiment that amounts to a direct demonstration of quantum realism (Long et al. 2018). The experiment is an extreme version of a "delayed-choice" experiment (Jacques et al. 2007). In Jacques' experiment, a single photon passes through BS1 and *then* a random choice is made to either insert or not insert BS2. Bohr's notions about the complementary dual wave-particle nature of quanta had misled some physicists into thinking a photon might mysteriously retain whichever aspect, wave or particle, it had acquired *before* entering the interferometer, regardless of the last-minute switching of BS2. As one might expect, this proved false. The photon interfered with itself if and only if BS2 was inserted, regardless of its history. A photon is not sometimes a wave and sometimes a particle; it is always a bundled lump of field that spreads whenever it can.

Realistically, a single photon is an EM wave packet stretched out in space. Accordingly, Long proposed to quickly insert (or remove) BS2 *while the photon is passing through the crossing*. Long's ingenious implementation allows him to insert BS2 at any time during the photon's passage through the crossing; his results show 16 data points corresponding to trials made at 16 "delay times" between 0 (when the photon's front end arrives at the crossing) and T (when its back end arrives). If Long had considered only delays of either 0 or T, the experiment would mimic Jacques' experiment, and indeed Long's results recapitulate Jacques' in these cases: phase-dependent interference with BS2 present (delay time of 0), phase-independent random 50-50 outcomes with BS2 absent (delay time of T).

For intermediate delay times, Long et al. analyze the four "sub-waves" (the two branches of the superposition, both before and after insertion of BS2) in detail to predict the probability of "particle-like" behavior (phase-independent random outcomes) and "wave-like" behavior (phase dependent interference) at the detectors, as a function of both the delay time and phase difference.

The experiment slices and dices each photon. The results for all four sub-waves are as expected from realistic standard QP, over all insertion times from 0 to T and all choices of phase between the two branches. It's hard to believe that a photon (and all other quanta) isn't just what a realist would have guessed decades ago: A real spatially-extended wave packet that spreads along all available paths.

**Nonlocality and Entanglement**

Localitt--the notion that objects are directly influenced only by their immediate surroundings--is intuitively plausible and is valid outside of QP. Long-distance forces such as classical gravity and electromagnetism act only "locally" because they travel as waves propagating from point to point, the changing gravitational or EM field at one point causing the field at the next point to change. But as we've seen, Planck's hypothesis implies that, if a single quantum has any spatial extension at all, then it must be *instantaneously* coordinated over its entire extent. Einstein noted this quantum wholeness in his 1927 remarks, as have others (Bohm 1980). Such wholeness is absent from classical physics but pervasive in QP and the source of quantum nonlocality.

Nonlocality is most obvious when two or more quanta are involved. When two quanta interact, their spatially extended states can become "entangled." Theory and experiment show entangled quanta can influence each other instantaneously regardless of their separation distance. Figure 3 indicates this conceptually. Assume the black quantum and gray quantum, initially moving rightward and upward respectively, are independent, i.e. alterations in one don't affect the other. They then come together and overlap their spatially extended states. This can "entangle" them so that, as they separate, each retains some portion of the other's state. The unity of each pre-entanglement quantum, implied by Planck's hypothesis, then suggests the entangled quanta form a single unified object.

Both the black and gray quanta now move in two ways simultaneously, in the following superposition:

> "Black quantum moves rightward and gray quantum moves upward," superposed with "black quantum moves upward and gray quantum moves rightward."

This two-quantum superposition turns out to be remarkable. As suggested by Figure 3, any alteration of one of the two final objects entails *simultaneous* changes in the other object, *regardless of their separation*. Such instantaneous coordination at a distance is perhaps not surprising when we recall the nonlocal implications of Planck's quantum hypothesis, and that quanta are simply ripples in a field: Two ripples have met and separated so that part of each incoming ripple now appears in both outgoing ripples.

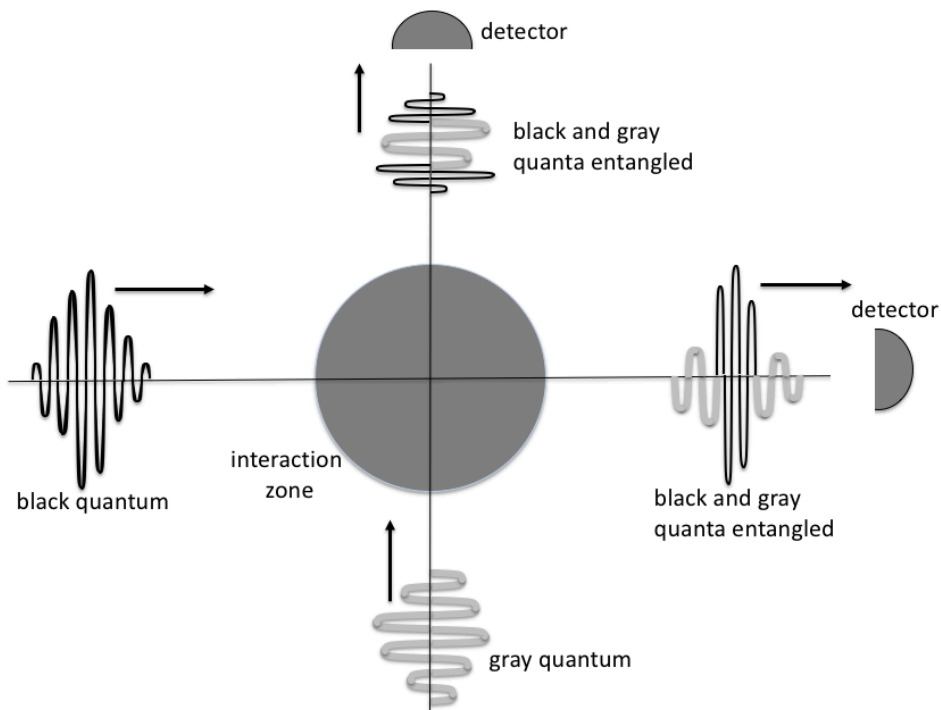

Figure 3.  Two quanta can entangle when they meet, a condition that persists even after separation.

Nonlocality is massively documented.  Einstein suggested nonlocal consequences of entanglement (Einstein, Podolsky & Rosen 1935); Bell found a mathematical criterion for whether entangled systems really behave nonlocally (Bell 1964); Clauser demonstrated nonlocality in entangled systems (Clauser & Freeman 1972); Aspect demonstrated nonlocal coordination occurs faster than lightspeed (Aspect, Dalibard & Roger 1982); and many experiments demonstrate nonlocality across great distances (Pan et al. 2017).  Nonlocality raised enormous skepticism, but experiments managed to close all loopholes simultaneously.[6]

The result is that it's now known beyond any reasonable doubt that "local realism" fails, meaning outcomes of measurements on entangled pairs of quanta are not fully determined by real properties (such as quantum states) carried along "locally" by either quantum.  Either individual ("local") systems are instantaneously influenced by distant events, or the properties (such as quantum states) of entangled systems don't objectively exist.  It's to quantum theory's great credit that it correctly predicts the failure of local realism.  But which one fails, locality or realism?  The Copenhagen camp claims realism fails, but others (Aspect 2007) argue locality fails.

It's helpful to consider a single experiment.  Of the plethora of nonlocality experiments since 1972, the most pedagogically useful are the experimental investigations of pairs of momentum-entangled photons performed nearly simultaneously by two independent groups (Rarity & Tapster 1990; Ou, Zou, Wang & Mandel 1990), referenced below as the "RTO experiments."  The results entirely agree with standard quantum theory (Horne, Shimony & Zeilinger 1990).

---

[6] See Giustina et al. (2015), Hensen et al. (2015), Shalm et al. (2015).

Figure 4 shows the layout. The source creates entangled photons A and B by a process that need not concern us. Note that A and B are the *post*-entanglement photons. Their entangled state can be described as follows:

"A moves leftward along the solid path and B moves rightward along the solid path," superposed with "A moves leftward along the dashed path and B moves rightward along the dashed path."

More briefly, A and B move away from each other along the solid path and also along the dashed path. A phase shifter is situated on each path.

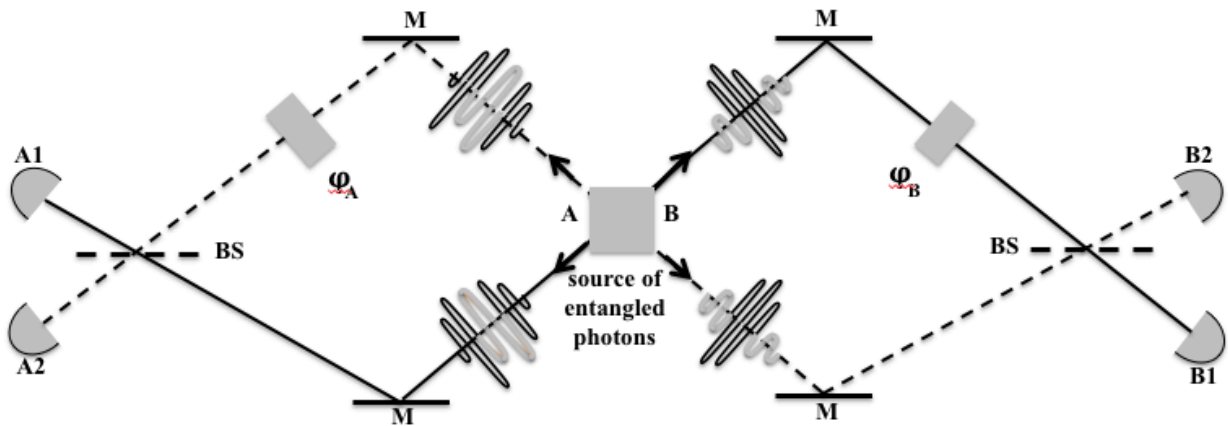

Figure 4. The RTO experiments.

If A and B were not entangled, the experiment would be simply two back-to-back interferometer experiments, each one like Figure 1 but with BS1 lying inside the source. However, the entanglement changes everything. The single photon of Figure 1 interferes only with itself, but neither photon in Figure 4 interferes with itself! This is a key result and as we'll see below it resolves the measurement problem. Regardless of either phase setting, both detectors (A1/A2 and B1/B2) register perfectly random 50-50 mixtures of outcomes. This means that neither photon has a phase of its own, even though non-entangled photons have phases of their own and can interfere with themselves.

However, the entangled pair *does* have a phase, and a state, of its own. Upon varying the phases we find the highly organized two-photon correlations graphed in Figure 5, showing both photons "know" the phase *difference*. The graph shows the photons' "degree of correlation" for various phase differences, where 360 degrees represents one wavelength difference between the solid and dashed path lengths. A "degree of correlation" of +1 means that, in a long series of trials, the two outcomes are always the same (either both are 1 or both are 2); a correlation of -1 means they are always different; zero correlation means the outcomes are the same on 50%, and different on 50%, of the trials; values between 0 and +1 indicate "same" is more likely than "different." For example, at a 45-degree phase difference the outcomes are the same in 71% of trials. For mathematical definitions, see (Hobson 2018).

*The perfect correlation at phase zero is manifestly nonlocal* because of the presence of the beam splitters: A's beam splitter sends photon A *randomly* to A1 or A2, and similarly for B's beam splitter. Nevertheless, each photon "knows" the "choice" of the other photon and correlates its path accordingly. How can A know B's choice? The answer: Entanglement. At zero phase, the two photons form a single object. In fact, it's enlightening to consider the *pair* of photons as a single "bi-photon," an "atom of light" (atoms are also internally entangled) superposed along two bi-photon paths, solid and dashed.

The coordination shown in Figure 5 is shocking, and entirely independent of the photons' separation: The entire bi-photon responds to changes in either phase shifter by adjusting its distant correlations accordingly. In fact, the experiment's violation of Bell's famous inequality shows that any change in phase entails an alteration in the future behavior of *both* photons. For example, if B's phase shifter changes, then both photons' future outcomes become different from what they would have been. Furthermore, the experiments of Aspect and others show this alteration occurs faster than light can travel between the two photons.

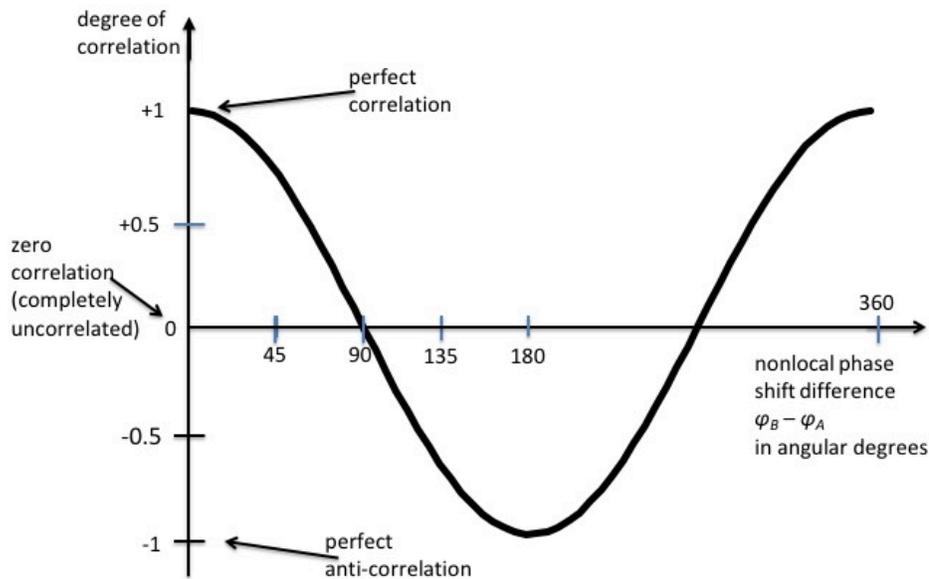

Figure 5. The degree of correlation between RTO's two entangled photons varies with the non-local phase difference $\phi_B - \phi_A$ between the photons.

Just as we can obtain Figure 2 by varying either phase shifter in Figure 1, we can obtain the correlations of Figure 5 by varying either phase shifter in Figure 4. Thus, just as Figure 2 entails that the single photon really is superposed along both paths, Figure 5 entails that the bi-photon really is superposed along both the solid path and the dashed path of Figure 4. The bi-photon simply spreads along both available paths, because it's a single unified field.

Table 1 compares outcomes for the simple superposition of Figure 1 and the entangled superposition of Figure 4, at five values of the phase difference from 0 to 180 degrees. The contrast is striking: In the simple superposition, the state of the single photon varies with phase, but the bi-photon's single-photon states show no such sign of either photon interfering with itself. Thus *the individual photons are no longer superposed, and only the correlations between the two photons vary with phase.* It's worth emphasizing that Figure 5 and Table 1 follow from standard QP (Horne, Shimony & Zeilinger 1990) and are verified by the RTO experiments.

Table I. Comparison of the simple superposition of Figure 1 with the entangled superposition of Figure 4.

| Simple superposition | | Entangled superposition | | |
|---|---|---|---|---|
| $\phi$ | State of photon | $\phi_B - \phi_A$ | State of each photon | Correlation between photons |
| 0 degrees | 100% 1, 0% 2 | 0 degrees | 50-50 1 or 2 | 100% corr, 0% anti |
| 45 | 71% 1, 29% 2 | 45 | 50-50 1 or 2 | 71% corr, 29% anti |
| 90 | 50% 1, 50% 2 | 90 | 50-50 1 or 2 | 50% corr, 50% anti |
| 135 | 29% 1, 71% 2 | 135 | 50-50 1 or 2 | 29% corr, 71% anti |
| 180 | 0% 1, 100% 2 | 180 | 50-50 1 or 2 | 0% corr, 100% anti |

It's been amply demonstrated that nonlocal correlations are established faster than light can connect the two measurements. This might seem to violate special relativity. But special relativity prohibits superluminal transfer only for causal signals. Nonlocality is implemented entirely with correlations, and correlations alone cannot transfer any causal effect. In the RTO experiment, for example, A cannot send an instantaneous signal to B by changing A's phase shifter, because nothing changes at B. The only way B can know something changed is by obtaining information about A's outcomes and noting the change of *correlations*. Thus, the bi-photon is instantaneously coordinated across its entire extent without violating special relativity.

We've seen that the RTO experiment is manifestly nonlocal. Nonlocality is also manifest in Planck's hypothesis, and in Einstein's comment at Solvay. Experiments such as RTO only reinforce the centrality of nonlocality to the quantum world. The consensus today is that either locality or realism fails, but there is no reason to abandon five centuries of scientific history by renouncing realism because nonlocality is clearly written all over QP.

A realistic picture of superposition and entanglement emerges. A fundamental quantum such as a photon is a highly unified bundle of field energy carrying one "excitation" (*hf* joules). In Figure 1, a photon spreads along paths 1 and 2. Upon measurement at D1/D2, the photon's reality emerges in its unified nature, its amplitudes (which needn't be 50-50), and its phase. Two entangled photons also form a single highly unified bundle of field energy but carrying two excitations. In Figure 4, a bi-photon spreads out along the solid and dashed paths and is detected at A1/A2 and B1/B2. Although the bi-photon is superposed, neither single photon is superposed. Neither photon has a phase (Hobson 2018), but the bi-photon has a phase and a quantum state as revealed by nonlocal correlations. Upon interaction, the bi-photon behaves as a single unified quantum regardless of its extension. A similar realistic view of superposition and entanglement has been expressed by Perez-Bergliaffa (Perez-Bergliaffa1996).

**Measurement and entanglement**

Measurements--processes in which a quantum event leads to an irreversible macroscopic change--have long presented difficulties for physicists. As will be shown, the main stumbling block has been entanglement. Entanglement is subtle. It is "*the* characteristic trait of quantum mechanics" (Schrodinger 1935, his emphasis). Today we still have no clear consensus about its nonlocal implications, although recent experiments might be changing this. Thus the continuing conundrum, and the variety of proposed solutions, are not surprising.

As a simple example, consider the single photon of Figure 1, without BS2, as the superposed photon moves along both paths toward D1/D2. As the photon nears the detector, the photon and detector interact to form the following entangled "measurement state":

"Photon on path 1 and D1 registers," superposed with "photon on path 2 and D2 registers."

Most physicists agree with the analysis up to this point, a point reached long ago in the first detailed analysis of measurement (von Neumann 1932).

The conundrum is that this state appears to entail that the macroscopic detector is superposed, and registers both D1 and D2. This would present two problems: (1) there is no definite outcome, so the measurement is useless; (2) such a macroscopic superposition, while physically possible, would be excruciatingly difficult to create and certainly could not be created as easily as this. Schrodinger dramatized the paradox in his famous example involving a radioactive nucleus, with a cat playing the role of detector by dying or living according to whether or not the nucleus decayed. The superposition seems to imply a dead and alive cat.

But we understand entanglement better today than in Schrodinger's day or indeed any other day, and this understanding resolves the supposed paradox (Hobson 2018). The above measurement state is simply a version of RTO's entangled state at zero phase (correlation of +1), with one of the quanta being (or triggering) a macroscopic detector. From Table 1, then, we see there is no paradoxical macroscopic superposition: the state of the photon is a random 50-50 probability of either path 1 or path 2, the state of the detectors is a random 50-50 probability of either D1 or D2, and the two outcomes are perfectly correlated. There is no macroscopic superposition because *neither subsystem is superposed*. In fact, we've seen that neither subsystem has a phase of its own, implying it cannot interfere with itself. Neither subsystem is a "coherent" quantum. Only *correlations between subsystems* are superposed, not *states* of subsystems, and thus there is no paradox. Even if one subsystem is macroscopic, there is no paradox. All this agrees with standard quantum theory (Horne, Shimony & Zeilinger 1990).

This solves the problem of "definite outcomes" also known as "Schrodinger's cat."

But it doesn't entirely solve the measurement problem, because a completed measurement must irreversibly record a single outcome. In fact, the mathematical formulation of the measurement process shows that the measurement state is in principle reversible, so it cannot yet be a permanent indicator such as a macroscopic sound wave or visible mark (Hobson 2018).

When the photon entangles with D1/D2 it transfers its energy to a single electron in either D1 or D2. This electron triggers a many-electron "avalanche," creating a detectable electric current in either D1 or D2. Experiments show the photon actually interacts with both detectors, but the triggering happens within just one of them while the other detector experiences only the quantum vacuum (Fuwa et al 2015). In other words, entanglement with *both* detectors ensures that *one* detector is triggered while the other is *not* triggered, resolving Einstein's question at the 1927 Solvay Conference. The process is necessarily non-local due to the separation between D1 and D2 and the requirement of instantaneous correlation. The indeterminism between D1 and D2 is a fundamental quantum property, just as it is throughout quantum physics.

Concerning irreversibility: the electron avalanche involves a large number of microscopic processes, each of them reversible or "unitary" in principle but collectively so numerous and diverse that the probability of reversal is essentially zero. Such processes are what the second law of thermodynamics is all about. The question of whether such a process is "really" irreversible or simply irreversible "for all practical purposes" has been argued since Ludwig Boltzmann's day (Hobson 1971), and arises in identical fashion even for classical systems. In other words, at this point our task of explaining quantum measurements is finished.

To summarize: Quantum measurement occurs when a superposed microscopic quantum entangles with a macroscopic object. The entanglement is not a paradoxical macroscopic superposition because (1) it is not a superposition of states of any single subsystem but merely a superposition of correlations between states of different subsystems; and (2) each subsystem has no phase and so cannot interfere with itself. The entanglement entails an indeterminate but definite quantum choice of macroscopic outcomes. The macroscopic object can be a laboratory detector or it can be natural, i.e. measurement has nothing necessarily to do with humans. Thus, when put into the proper realistic physical context, the measurement problem is not difficult.

**Conclusion**

Properly understood, and as has also been argued by Mario Bunge, the issues of quantization, field-particle duality, superposition, entanglement, nonlocality, and measurement present no barrier to a consistent and realistic interpretation based on standard quantum physics. At least to this extent, quantum physics is consistent with the scientific view as it has been known since Copernicus: nature exists on its own and science's goal is to understand its operating principles, which are independent of humans.